\documentclass[aps,prc,twocolumn,superscriptaddress]{revtex4-1}
\usepackage{graphicx}

\begin{document}


\title{First observation of excited states of $^{173}$Hg}


\author{D.~O'Donnell}
\email{david.odonnell@liv.ac.uk}
\affiliation{Oliver Lodge Laboratory, University of Liverpool, Liverpool, L69 7ZE, United Kingdom.}

\author{R.D.~Page}
\affiliation{Oliver Lodge Laboratory, University of Liverpool, Liverpool, L69 7ZE, United Kingdom.}

\author{C.~Scholey}
\affiliation{Department of Physics, University of Jyv\"askyl\"a, PO Box 35, FI-40014, Jyv\"askyl\"a, Finland.}

\author{L.~Bianco}
\affiliation{Department of Physics, University of Guelph, Guelph, Ontario, N1G 2W1, Canada.}

\author{L.~Capponi}
\affiliation{SUPA, School of Engineering, University of the West of Scotland,
High Street, Paisley, PA1 2BE, United Kingdom.}

\author{R.J.~Carroll}
\affiliation{Oliver Lodge Laboratory, University of Liverpool,
Liverpool, L69 7ZE, United Kingdom.}

\author{I.G.~Darby}
\altaffiliation[Present address: ]{IAEA Nuclear Spectrometry and Applications Laboratory, Physics Section, A-2444, Siebersdorf, Austria.}
\affiliation{Department of Physics, University of Jyv\"askyl\"a, PO Box 35, FI-40014, Jyv\"askyl\"a, Finland.}

\author{L.~Donosa}
\affiliation{Oliver Lodge Laboratory, University of Liverpool,
Liverpool, L69 7ZE, United Kingdom.}

\author{M.~Drummond}
\affiliation{Oliver Lodge Laboratory, University of Liverpool,
Liverpool, L69 7ZE, United Kingdom.}

\author{F.~Ertu\u{g}ral}
\affiliation{Physics Department, Faculty of Arts and Sciences,
Sakarya University, 54100 Serdivan, Adapazari, Turkey.}

\author{T.~Grahn}
\affiliation{Department of Physics, University of Jyv\"askyl\"a,
PO Box 35, FI-40014, Jyv\"askyl\"a, Finland.}

\author{P.T.~Greenlees}
\affiliation{Department of Physics, University of Jyv\"askyl\"a,
PO Box 35, FI-40014, Jyv\"askyl\"a, Finland.}

\author{K.~Hauschild}
\affiliation{Department of Physics, University of Jyv\"askyl\"a,
PO Box 35, FI-40014, Jyv\"askyl\"a, Finland.}
\affiliation{CSNSM, IN2P3-CNRS et Universit\'{e} Paris Sud,
Paris, France}

\author{A.~Herzan}
\affiliation{Department of Physics, University of Jyv\"askyl\"a,
PO Box 35, FI-40014, Jyv\"askyl\"a, Finland.}

\author{U.~Jakobsson}
\affiliation{Department of Physics, University of Jyv\"askyl\"a,
PO Box 35, FI-40014, Jyv\"askyl\"a, Finland.}

\author{P.~Jones}
\affiliation{Department of Physics, University of Jyv\"askyl\"a,
PO Box 35, FI-40014, Jyv\"askyl\"a, Finland.}

\author{D.T.~Joss}
\affiliation{Oliver Lodge Laboratory, University of Liverpool,
Liverpool, L69 7ZE, United Kingdom.}

\author{R.~Julin}
\affiliation{Department of Physics, University of Jyv\"askyl\"a,
PO Box 35, FI-40014, Jyv\"askyl\"a, Finland.}

\author{S.~Juutinen}
\affiliation{Department of Physics, University of Jyv\"askyl\"a,
PO Box 35, FI-40014, Jyv\"askyl\"a, Finland.}

\author{S.~Ketelhut}
\affiliation{Department of Physics, University of Jyv\"askyl\"a,
PO Box 35, FI-40014, Jyv\"askyl\"a, Finland.}

\author{M.~Labiche}
\affiliation{STFC Daresbury Laboratory, Daresbury, Warrington,
WA4 4AD, United Kingdom.}

\author{M.~Leino}
\affiliation{Department of Physics, University of Jyv\"askyl\"a,
PO Box 35, FI-40014, Jyv\"askyl\"a, Finland.}

\author{A.~Lopez-Martens}
\affiliation{Department of Physics, University of Jyv\"askyl\"a,
PO Box 35, FI-40014, Jyv\"askyl\"a, Finland.}
\affiliation{CSNSM, IN2P3-CNRS et Universit\'{e} Paris Sud,
Paris, France}

\author{K.~Mulholland}
\affiliation{SUPA, School of Engineering, University of the West of Scotland,
High Street, Paisley, PA1 2BE, United Kingdom.}

\author{P.~Nieminen}
\affiliation{Department of Physics, University of Jyv\"askyl\"a,
PO Box 35, FI-40014, Jyv\"askyl\"a, Finland.}

\author{P.~Peura}
\affiliation{Department of Physics, University of Jyv\"askyl\"a,
PO Box 35, FI-40014, Jyv\"askyl\"a, Finland.}

\author{P.~Rahkila}
\affiliation{Department of Physics, University of Jyv\"askyl\"a,
PO Box 35, FI-40014, Jyv\"askyl\"a, Finland.}

\author{S.~Rinta-Antila}
\affiliation{Department of Physics, University of Jyv\"askyl\"a,
PO Box 35, FI-40014, Jyv\"askyl\"a, Finland.}

\author{P.~Ruotsalainen}
\affiliation{Department of Physics, University of Jyv\"askyl\"a,
PO Box 35, FI-40014, Jyv\"askyl\"a, Finland.}

\author{M.~Sandzelius}
\affiliation{Department of Physics, University of Jyv\"askyl\"a,
PO Box 35, FI-40014, Jyv\"askyl\"a, Finland.}

\author{J.~Sar\'en}
\affiliation{Department of Physics, University of Jyv\"askyl\"a,
PO Box 35, FI-40014, Jyv\"askyl\"a, Finland.}

\author{B.~Saygi}
\affiliation{Oliver Lodge Laboratory, University of Liverpool,
Liverpool, L69 7ZE, United Kingdom.}

\author{J.~Simpson}
\affiliation{STFC Daresbury Laboratory, Daresbury, Warrington,
WA4 4AD, United Kingdom.}

\author{J.~Sorri}
\affiliation{Department of Physics, University of Jyv\"askyl\"a,
PO Box 35, FI-40014, Jyv\"askyl\"a, Finland.}

\author{A.~Thornthwaite}
\affiliation{Oliver Lodge Laboratory, University of Liverpool,
Liverpool, L69 7ZE, United Kingdom.}

\author{J.~Uusitalo}
\affiliation{Department of Physics, University of Jyv\"askyl\"a,
PO Box 35, FI-40014, Jyv\"askyl\"a, Finland.}


\date{\today}

\begin{abstract}
The neutron-deficient nucleus $^{173}$Hg has been studied following fusion-evaporation reactions. The observation of the decay of excited states via $\gamma$ radiation are reported for the first time and a tentative level scheme is proposed. The proposed level scheme is discussed within the context of the systematics of neighboring neutron-deficient Hg nuclei. In addition to the $\gamma$-ray spectroscopy, the $\alpha$ decay of this nucleus has been measured yielding superior precision to earlier measurements.
\end{abstract}


\maketitle


\section{Introduction}
The neutron-deficient Hg atomic nuclei have attracted much attention in recent years. In-beam studies of even-even
isotopes~\cite{Dracoulis_180Hg,Kondev_178Hg,Muikku_176Hg,Uusitalo_174Hg,Sandzelius_172Hg} have allowed the evolution of the nuclear shape to be
investigated right to the proton drip line~\cite{Sandzelius_172Hg} and show a clear trend of the ground state becoming increasingly spherical with
decreasing neutron number. This is in stark contrast to the shape coexistence observed near the neutron mid-shell~\cite{Julin_intruders}. In
addition, the odd-$A$ Hg isotopes have been the subject of scrutiny during the last
decade~\cite{Kondev_179Hg,Jenkins_179Hg,Melerangi_177Hg,ODonnell_175Hg} with detailed spectroscopy being performed down to
$^{175}$Hg~\cite{ODonnell_175Hg}. The study of the odd-$A$ isotopes has proved useful in helping to identify the active single-particle orbitals
driving the changes in shape observed in the neighboring even-$A$ nuclei. In particular, the observed increase in excitation energy of the $J^\pi =
13/2^+$ $\nu i_{13/2}$ band-head from $^{179}$Hg~\cite{Kondev_179Hg,Jenkins_179Hg} to $^{175}$Hg~\cite{ODonnell_175Hg} is indicative of the
diminishing influence of this deformation-driving orbital on the shape of the neighboring even-$A$ nuclei.

The rapid progress that has been made in the studies of the lightest Hg isotopes belies the experimental challenges faced when attempting to study such neutron-deficient nuclei. These nuclei tend to be populated via the evaporation of several neutrons following fusion-evaporation reactions and the cross sections for their production are low when compared with competing reaction channels, not to mention the competition introduced by fission of the compound system. The technique of recoil-decay tagging (RDT)~\cite{Schmidt_RDT,Simon_RDT,Paul_RDT}, in which a highly efficient $\gamma$-ray spectrometer is coupled to a recoil separator with a detection system located at the focal plane, has enabled these  difficulties to be overcome and permitted detailed spectroscopy of many neutron-deficient nuclei, including the Hg isotopes.

The $^{173}$Hg isotope was first reported by Seweryniak {\it et al.}~\cite{Seweryniak_173Hg} who measured the energy (7211(11)~keV) and half-life ($0.93^{+0.57}_{-0.26}$~ms) of the $\alpha$ decay from the observation of only seven events. The only other study of the properties of $^{173}$Hg was the work of Kettunen {\it et al.}~\cite{Kettunen} in which an energy of 7192(13)~keV and a half-life of $0.59^{+0.47}_{-0.18}$~ms were reported. In the present work the RDT technique has been utilized in order to observe excited states of $^{173}$Hg for the first time. A level scheme representing the decay of these states is proposed and is discussed within the context of the behavior of excited states of neighboring Hg nuclei. In addition, the previous measurements of the $\alpha$-decay energy and half-life have been improved upon.

\section{Experimental details}
The $^{173}$Hg nuclei were produced following the complete fusion of $^{84}$Sr and $^{92}$Mo and the subsequent evaporation of three neutrons from the $^{176}$Hg compound nucleus. The $^{84}$Sr$^{16+}$ ions were accelerated by the K130 cyclotron of the University of Jyv\"{a}skyl\"{a}. The beam was incident on the target for a total of 285 hours. For approximately half of this time (140 hours) the beam was provided with an energy of 392~MeV and for the remaining 145 hours the energy was increased to 400~MeV. An average beam current of $\approx$ 120~enA was incident on the 600~$\mu$g/cm$^{2}$ $^{92}$Mo target which was isotopically enriched to 98~\%.

The target position was surrounded by 34 large-volume, high-purity Ge detectors in order to measure the energy of prompt $\gamma$ radiation following the decay of excited states of the fusion-evaporation residues (recoils). These detectors consisted of two different types: 10 EUROGAM Phase-I detectors~\cite{Beausang_Jurogam} positioned at 133.6$^\circ$ with respect to the beam direction and 24 Clover-type detectors~\cite{Duchene_clover} at 104.5$^\circ$ and 75.5$^\circ$. The RITU gas-filled magnetic separator~\cite{Leino1} was used to separate the recoils, which were travelling with a velocity of approximately 0.045$c$, from scattered beam and transport the former to the focal plane.

The GREAT spectrometer~\cite{Page_GREAT}, which consisted of an isobutane-filled multi-wire proportional chamber (MWPC), two adjacent double-sided silicon strip detectors (DSSDs), a planar Ge detector, an array of Si PIN diodes and four large-volume Ge detectors, was located at the RITU focal plane in order to detect delayed radiation associated with the decay of the reaction products. Time-of-flight measurements, corresponding to the difference in time between detection of recoils in the MWPC and DSSDs, in addition to energy-loss information from the MWPC, allowed the discrimination of recoils and any remaining scattered beam.

The reaction products were subsequently implanted into the highly-segmented (4800 pixels) DSSDs; the front side of which was instrumented to detect
electrons produced as a result of internal conversion while the rear side was set up to detect $\alpha$ particles. The array of Si PIN diode
detectors was also used to detect conversion electrons while the Ge detectors were used to detect $\gamma$ rays. The Total Data Readout
system~\cite{TDR} was employed in order to record signals observed in each of the constituent detectors, time stamped with a precision of 10~ns. In
this way it was possible in the offline analysis to correlate events with a characteristic $\alpha$ decay and unambiguously assign any prompt and
delayed $\gamma$ rays or conversion electrons to the decay of a specific nucleus. The data were sorted and analyzed offline using the Grain software
package~\cite{Rahkila_Grain}.

\begin{figure}
\includegraphics[scale=0.34,angle=270,clip]{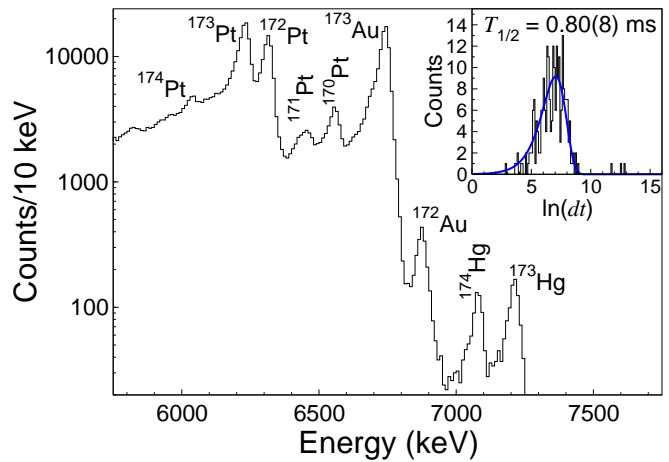}
\caption{\label{alphas} (Color online) Spectrum showing all $\alpha$ particles observed within 7~ms of the implantation of a recoil in the same pixel of the DSSD. The time distribution of detected $\alpha$ particles identified with the decay of implanted $^{173}$Hg nuclei which have been correlated with their daughter activity ($^{169}$Pt) is shown in the inset. The full blue line represents a fit to the data.}
\end{figure}

\section{Experimental results}
In Figure~\ref{alphas} is shown events detected in the DSSDs within 7~ms of the implantation of a fusion-evaporation residue in the same pixel. The DSSDs were calibrated using $\alpha$ lines at 5407, 6038 and 6315~keV corresponding to the known activities of $^{170}$Os~\cite{A166_2008} (not visible in Figure~\ref{alphas}), $^{174}$Pt~\cite{A170_2002} and $^{172}$Pt~\cite{A168_2010}, respectively. The $\alpha$ particles associated with the decay of $^{173}$Hg were measured to have an energy of 7208(5)~keV which is in agreement with the previously reported values~\cite{Seweryniak_173Hg,Kettunen} but represents a more precise measurement. In total, approximately 1000 recoil-$\alpha$($^{173}$Hg) events were observed throughout the $\approx$285 hours of the experiment. Assuming an efficiency of 40\% for the transportation of the $^{173}$Hg nuclei through RITU~\cite{Saren_RITU}, the cross section for the production of $^{173}$Hg in the present work is estimated to be 30~nb.

In the inset to Figure~\ref{alphas} is shown the time distribution of $\alpha$ particles, consistent with the known decay properties of $^{173}$Hg, detected within 500~ms of an implanted recoil and which were followed by $\alpha$ decays of $^{169}$Pt ($E = 6695(5)$~keV, $T_{1/2} = 6.99(10)$~ms~\cite{ODonnell_168Pt}) in the same pixel. The form of this plot follows the method outlined by Schmidt {\it et al.}~\cite{Schmidt_max_likelihood} in which the difference in time between implantation and decay ($dt$) are plotted on a logarithmic scale. The most likely value of the distribution shown in the inset to Fig.~\ref{alphas} is then equal to ln$(\tau)$ where $\tau$ is the lifetime of the $\alpha$-decaying state. The method of maximum likelihood~\cite{Schmidt_max_likelihood} has yielded a half life for the $\alpha$ decay of $^{173}$Hg of 0.80(8)~ms. Again, this is in good agreement with the previously reported values~\cite{Seweryniak_173Hg,Kettunen} but is a more precise measurement. Using the $\beta$-decay half-life predicted by Moller {\it et al.}~\cite{MollerNixKratz}, the $\alpha$-decay branching ratio of $^{173}$Hg is estimated to be 100(4)\% which yields a reduced width of 80(9)~keV indicating a likely $\Delta L =0$ transition to the $^{169}$Pt daughter. Although the spin and parity of the ground state of $^{169}$Pt has, to date, not been measured, it is plausible that it has $J^\pi = 7/2^-$ since the $^{169}$Pt-$^{165}$Os-$^{161}$W-$^{157}$Hf $\alpha$ decay chain proceeds to the ground state of $^{157}$Hf ($J^\pi = 7/2^-$~\cite{Saad_157Hf}) via transitions consistent with no change in angular momentum~\cite{Page_radioactivity_survey}. This argument assumes ground state to ground state transitions for each stage of this decay chain.

\begin{figure}
\includegraphics[scale=0.34,angle=270,clip]{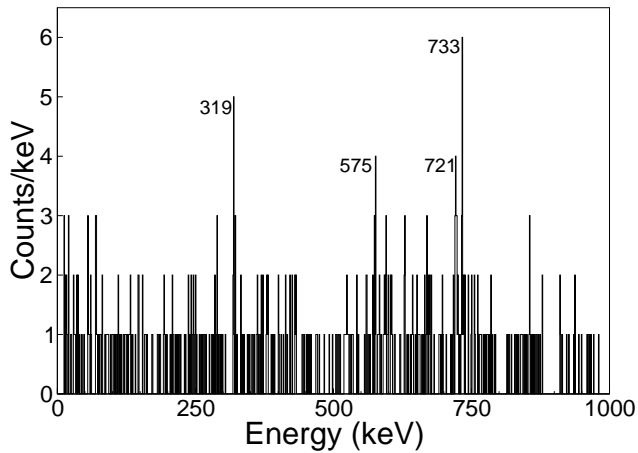}
\caption{\label{gammas} Prompt $\gamma$ rays detected at the target position in delayed coincidence with recoils implanted into a DSSD and followed within 7~ms by an $\alpha$ decay of $^{173}$Hg in the same pixel.}
\end{figure}

\begin{figure}
\includegraphics[scale=0.34,angle=270,clip]{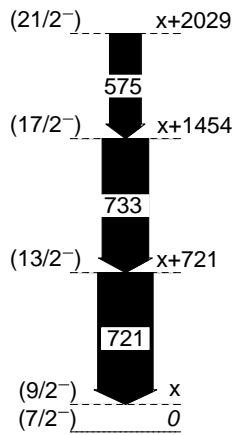}
\caption{\label{level_scheme} The proposed tentative level scheme representing the decay of excited states of $^{173}$Hg as observed in the present work. The width of the arrows are proportional to the measured intensity of the transitions.}
\end{figure}

In Figure~\ref{gammas} is shown the prompt $\gamma$ radiation, detected at the target position, associated with recoils implanted into the DSSDs at the focal plane and which were followed within 7~ms of a $^{173}$Hg $\alpha$ decay. Four transitions have been identified and, as a result, unambiguously associated with the decay of excited states of $^{173}$Hg. The energy of these transitions are (in order of decreasing intensity): 721 ($I_\gamma = 100(33)$), 733 ($I_\gamma = 84(31)$), 575 ($I_\gamma = 57(21)$) and 319~keV ($I_\gamma = 30(13)$). The low $\gamma$-ray statistics obtained have prevented a $\gamma\gamma$ coincidence analysis therefore rendering the construction of a level scheme difficult. Nonetheless, a tentative level scheme is proposed and shown in Figure~\ref{level_scheme}. It should be noted that the uncertainties on the intensity measurements for the 721 and 733~keV transitions indicate that the ordering of these transitions could be reversed. For the purposes of building the level scheme it has been assumed that the transitions included in the level scheme are stretched quadrupole in nature and form a cascade.

\section{Discussion}
\begin{figure}
\includegraphics[scale=0.44,angle=0,clip]{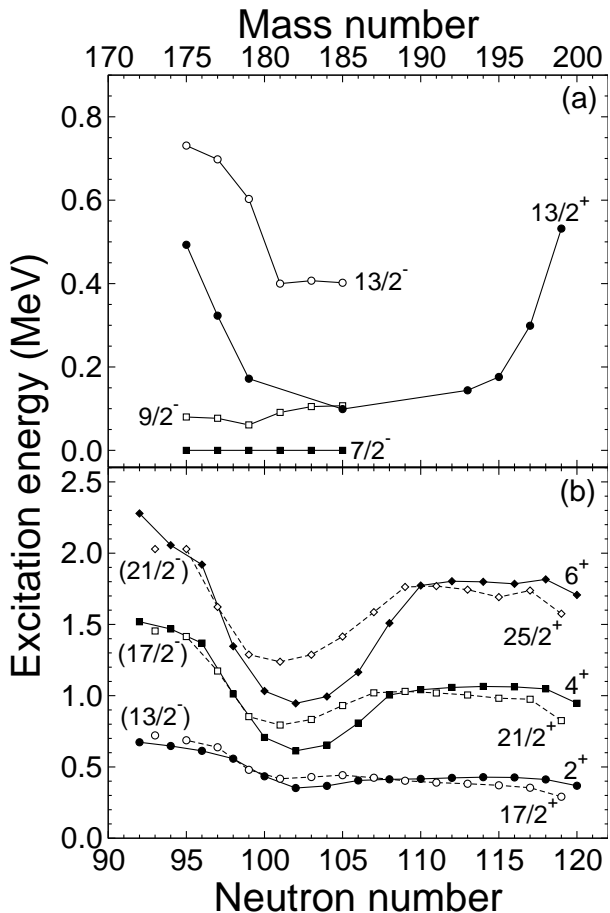}
\caption{\label{systematics} Plot showing (a) the excitation energies of four states common to a number of odd-$A$ $_{80}$Hg isotopes; and (b) excitation energies of the first excited $J^\pi = 2^+, 4^+$ and $6^+$ states in even-$A$ (filled symbols) and the $J^\pi = 17/2^+, 21/2^+$ and $25/2^+$ for the odd-$A$ (open symbols) Hg isotopes with the exception of $^{173}$Hg where the excited states proposed here have been plotted. The odd-$A$ energies have been normalized with respect to the energy of the band head.}
\end{figure}

The lightest odd-$A$ Hg isotopes ($A \leq 179$) have been reported as having $J^\pi = 7/2^-$ ground states which have been attributed to a mixed configuration involving the $f_{7/2}$ and $h_{9/2}$ neutron orbitals~\cite{Jenkins_179Hg,Melerangi_177Hg,ODonnell_175Hg}. Low-lying structures based on these configurations are also reported in $^{181-185}$Hg but do not constitute the ground state in these nuclei~\cite{Varmette_181Hg,Lane_183Hg,Hannachi_185Hg} which have $J^\pi = 1/2^-$ in each case. In Figure~\ref{systematics}(a) is shown the excitation energies, as a function of neutron number, for a few of the low-lying states common to many odd-$A$ Hg nuclei. In particular, three negative parity states have been plotted, their energies normalized to the energy of the $7/2^-$ level while the unnormalized energies of the $13/2^+$ states have also been plotted. One noteworthy feature of this plot is how relatively constant the $9/2^- \rightarrow 7/2^-$ transition is observed to be, remaining within 46~keV across six nuclides. An extrapolation of this trend to $N = 93$ would suggest that the ground state of $^{173}$Hg has $J^\pi = 7/2^-$ and that the $9/2^-$ state in $^{173}$Hg would be found between 80 and 100~keV higher in excitation energy. As can be seen from Figure~\ref{gammas}, no prompt $\gamma$-ray transition was observed in this energy range. However, such a low energy M1 transition would have a large internal conversion coefficient ($\alpha \geq 2.5$~\cite{BrIcc}) and, based on the observed $\gamma$-ray statistics, would not be discernable from background. Therefore, since the $9/2^-$ state is expected to lie at an unknown excitation energy above the ground state it has been included in the level scheme of Figure~\ref{level_scheme}.

It is possible, given the behavior of the systematics plotted in Figure~\ref{systematics}(a), that the $13/2^-$ state becomes yrast at $N = 93$. This state would likely be the result of a coupling of the $h_{9/2}$ neutron to the $2^+$ excitation of the core. The energies of excited states in the Hg isotopes have been plotted as a function of neutron number in Figure~\ref{systematics}(b). In the case of the even-$A$ isotopes (filled symbols) the excitation energies of the yrast $J^\pi = 2^+, 4^+ \text{and}~6^+$ states are plotted. In all but one ($^{173}$Hg) of the odd-$A$ isotopes it is the yrast $J^\pi = 17/2^+, 21/2^+ \text{and}~25/2^+$ excitation energies, relative to the $13/2^+$ energy, which have been plotted. It can be seen from Figure~\ref{systematics}(b) that the energies of the proposed excited states of $^{173}$Hg fit reasonably well within the context of the neighboring even- and odd-$A$ isotopes. In particular, there is very little change in excitation energies of the yrast states in moving from $^{175}$Hg to $^{173}$Hg, indeed it would seem that $^{173}$Hg could be best described as a neutron-hole coupled to the $^{174}$Hg core. As a result, it is assumed that the higher-lying excited states of $^{173}$Hg also result from the weak coupling of the odd neutron and core excitations.

In Figure~\ref{systematics}(a) the parabolic-like behavior of the $13/2^+$ states with varying neutron number can be seen. These states have been associated with the $\nu 13/2[606]$ Nilsson orbital and, for $N \leq 99$, have been observed to be isomeric, decaying via M2 transitions ($T_{1/2} = 0.34~\mu$s, $1.5~\mu$s and $6.4~\mu$s for $^{175,177,179}$Hg, respectively) to the $9/2^-$ states~\cite{Jenkins_179Hg,Melerangi_177Hg,ODonnell_175Hg}. If the parabolic-like trend is extrapolated to $N = 93$ then a $13/2^+$ state could be expected to have an excitation energy in excess of 850~keV in $^{173}$Hg. At such an energy, it is estimated that an M2 transition to a $9/2^-$ state would have a half-life of the order of 5~ns. Such a transition is unlikely to be observed in the current work given the level of statistics.

No evidence of an isomeric transition was found for $^{173}$Hg as a result of the present work, despite the five Ge detectors employed at the focal plane of RITU. The non-observation of any delayed transitions at the focal plane is in-line with expectations that the excitation energy of the $13/2^+$ state has increased to the extent that it is no longer yrast. As this state results from the occupation of the $\nu i_{13/2}$ orbital the behavior of the $13/2^+$ level is indicative of the influence which the deformation-driving single-particle state has on the low-lying structure of the neighboring even-$A$ Hg isotopes. The observed increase in energy for $A \leq 179$ and the fact that it is no longer yrast in $^{173}$Hg goes some way to explain the increasing sphericity reported in even-$A$ isotopes.

In terms of the future and supplementing the present work, it would be interesting to employ an efficient and high-resolution electron spectrometer at the target position. This could help to determine the existence and the excitation energy of the expected $9/2^-$ first excited state by measuring the electrons emitted as a result of the expected M1 transition to the ground state. Independently of these measurements there is scope for further investigations of this nucleus in the form of the $\alpha$ decay of the as yet unknown nuclide, $^{177}$Pb, which should be achievable with current apparatus. The major limiting factor with such a study would be the availability of appropriate beam and target materials.

\begin{acknowledgments}
Financial support for this work has been provided by the UK Science and Technology Facilities Council (STFC) and by the EU 7$^{th}$ framework
programme "Integrating Activities - Transnational Access", project number: 262010 (ENSAR) and by the Academy of Finland under the Finnish Centre of
Excellence Programme 2012-2017 (Nuclear and Accelerator Based Physics Research at JYFL). The authors would like to express their gratitude to the
technical staff of the Accelerator Laboratory at the University of Jyv\"askyl\"a for their support.
\end{acknowledgments}

\end{document}